\documentclass{elsart}
\journal{International Journal of Modern Physics B}

\usepackage{graphicx,subfigure,color}

\usepackage{amssymb,amsmath}
\usepackage{cite}
\usepackage{url}

\definecolor{dgreen}{rgb}{0,.6,0}

\usepackage[normalem]{ulem}

\begin{document}

\begin{frontmatter}

\title{Cryptanalysis of a new chaotic cryptosystem based on ergodicity}
\author[Spain]{David Arroyo\corauthref{corr}},
\author[Spain]{Gonzalo Alvarez},
\author[germany]{Shujun Li},
\author[hk-cityu]{Chengqing Li} and
\author[Spain]{Veronica Fernandez}
\corauth[corr]{Corresponding author: David Arroyo
(david.arroyo@iec.csic.es).}
\address[Spain]{Instituto de F\'{\i}sica Aplicada, Consejo Superior de
Investigaciones Cient\'{\i}ficas, Serrano 144, 28006 Madrid, Spain}
\address[germany]{FernUniversit\"{a}t in Hagen, Chair of Computer Engineering, Universit\"{a}tsstra{\ss}e 27, 58084 Hagen, Germany}
\address[hk-cityu]{Department of Electronic Engineering, City University of Hong Kong,
83 Tat Chee Avenue, Kowloon Tong, Hong Kong SAR, China}

\begin{abstract}
This paper analyzes the security of a recent cryptosystem based on
the ergodicity property of chaotic maps. It is shown how to obtain
the secret key using a chosen-ciphertext attack. Some other design
weaknesses are also shown.
\end{abstract}

\end{frontmatter}

\section{Introduction}
Chaotic maps possess an ergodic behavior which makes them suitable
for the design of new cryptosystems. This is the case of the
cryptosystem proposed in \cite{alvarez99}. This cryptosystem is
based on the tent map and has been cryptanalyzed in \cite{alvarez00}
and later improved in \cite{li2001}. In \cite{wang08b} a new
modification on the original scheme described in \cite{alvarez99}
was proposed. The authors of this new proposal claim that this
modification overcomes all the security problems that were
emphasized in \cite{alvarez00,li2001}. Nevertheless, in this paper
we show that the ciphertext still includes enough information to
enable a chosen-ciphertext attack based on symbolic dynamics. The
rest of the paper is organized as follows. First of all,
Sec.~\ref{section:description} gives a brief introduction to the
cryptosystem under study. After that, in Sec.~\ref{section:chosen}
the symbolic dynamics based chosen-ciphertext attack is explained.
Then some other problems of the cryptosystem under study are
discussed in Sec.~\ref{section:weaknesses}, and finally the last
section gives some final comments and conclusions.

\section{Description of the cryptosystem}
\label{section:description}

The cryptosystem described in \cite{wang08b} is based on the
transformation of chaotic orbits into binary sequences. These
chaotic orbits are generated using a one-dimensional chaotic map
defined by
\begin{equation}
    x_{n+1} = f(x_n,r),
    \label{eq:iteration}
\end{equation}
where $f: \mathcal{I} \rightarrow \mathcal{I}$ and
$0.5\in\mathcal{I}\subset\mathbb{R}$. If Eq.~\eqref{eq:iteration} is
iterated $N$ times, then a chaotic orbit will be obtained as
\begin{equation}
    \left\{x_n\right\}_{n=0}^{N}=
    \left\{x_0,x_1,\ldots,x_{N}\right\}.
    \label{eq:orbit}
\end{equation}
The authors of \cite{wang08b} do not explicitly indicate if $x_0$ is
also included in the chaotic orbit as the first chaotic state.
Without loss of generality, in this paper we will assume that this
was included.

Finally, the binary counterpart (i.e., the symbolic dynamics based
representation) of the original chaotic orbit is given by
\begin{equation}
    g_n = g_n(x_0,r)=\left\{ \begin{array}{ll}0 &\textrm{, if} \ \ x_n<0.5,\ \ \ \ \ \ \\
    1 & \textrm{, if} \ \ x_n \geq 0.5 ,\end{array}\right.
    \label{eq:gray}
\end{equation}
for $0\leq n \leq N$. Henceforth, the binary sequence
$\left\{g_n(x_0,r)\right\}_{n=0}^{N}$ is noted as $G^N(x_0,r)$ to
emphasize its dependency with the initial condition and the control
parameter.

The cryptosystem works as follows.

\begin{itemize}

\item
Step 1) Initialize $i=0, j=0$.

\item
Step 2) For the $i$-th plain block $P_i$ formed by $b_i=b$ bits, try
to find the first $b_i$-bit segment of $\{g_n\}_{n=j}^{N_{max}+b_i}$
which is equal to $P_i$; in case a segment is not found, let
$b_i=b_i-1$ and repeat this step\footnote{Note that in
\cite{wang08b}, there was a typo about $b_i=b_i-1$, which was
published as ``$b_i=b_i+1$''.}. The parameter $N_{max}$ indicates
the maximum number of trials in the searching of $P_i$ through the
binary sequence.

\item
Step 3) Denoting by $n_i$ the number of iterations needed to locate
the distinguished $b_i$-bit segment from $g_j$, output $(b_i,n_i)$
as the $i$-th cipher-block.

\item
Step 4) Set $i=i+1$ and $j=j+n_i+b_i$ \footnote{In \cite{wang08b} it
is not explicitly mentioned how to update the index $j$. In this
paper, we assume that it is updated in such a way that no segment of
a chaotic orbit will be reused for encryption of two continuous
plain blocks.}, then go to Step 2 until the whole plaintext is
exhausted.

\end{itemize}

The decryption process is simpler than the encryption one. In this
case, the searching process becomes unnecessary. For the recovery of
the $i$-th plain block, one simply iterates the chaotic map from the
current status for $n_i+b_i$ times and record the last $b_i$ chaotic
states, which are then transformed into the $i$-th $b_i$-bit plain
block according to Eq.~\eqref{eq:gray}.

In \cite{wang08b} it is claimed that the secret key of the
cryptosystem is composed of the initial condition $x_0$ and the
control parameter $r$. For a more detailed description of the
encryption/decryption procedures, the reader is referred to
\cite{wang08b}.

\section{Chosen-ciphertext attack}
\label{section:chosen}

In \cite{wang08b} it is mentioned that most chaotic systems can be
used to implement the above described cryptosystem. Moreover, the
resistance of the cryptosystem against the attacks presented in
\cite{alvarez00} is assumed without any security analysis. However,
this section proves that a wrong selection of the chaotic map allows
an estimation of the secret key through a chosen-ciphertext attack.

Among all the possible options, the logistic map was chosen in
\cite{wang08b} as the chaotic system to prove the reliability of the
cryptosystem. The logistic map is defined as
\begin{equation}
    x_{n+1}= f(x_n,r) = r \cdot x_n\cdot (1-x_n),
    \label{eq:logistic}
\end{equation}

for $r \in (3.57148, 4)$ and $x_n \in [0,1]$. The function $f(x,r)$
for the logistic map is a concave function with only one critical
point at $0.5$. For this kind of maps the binary sequence referred
in Eq.~\eqref{eq:gray} can be interpreted as a Gray code
\cite{alvarez98,cusick99}. Moreover, in
\cite{alvarez03a,wu04,physcon07} it is shown that the family of Gray
codes generated using Eq.~\eqref{eq:gray} can be assigned an order
according to the initial condition and the control parameter. The
existence of this order allows an estimation of the control
parameter $r$ and the initial condition $x_0$ just by analyzing the
binary sequence $G^N(x_0,r)$ for a sufficiently large number $N$.
Therefore, as long as one can reconstruct the sequence $G^N(x_0,r)$,
one can estimate the secret key of the cryptosystem. This is used to
build an attack with three different stages:

\begin{enumerate}
    \item Reconstruction of the Gray code derived from the logistic
    map.
    \item Estimation of the control parameter from the
    reconstructed Gray code.
    \item Estimation of the initial condition from the reconstructed
    Gray code and the estimated control parameter.
\end{enumerate}

\subsection{Reconstruction of the Gray code}

If one has access to the decryption machine, then one can perform a
chosen-ciphertext attack \cite[p. 25]{stinson95} to reconstruct
$G^N(x_0,r)$, i.e., the Gray code associated to the values of $x_0$
and $r$ that make up the secret key of the cryptosystem under study.
To do so, $M$ ciphertexts are generated as $(b,b\cdot i)$ for
$i=0,1,2,\ldots,M$. As an example, let us assume that $x_0=0.5$ and
$r=3.78$. In this case, it is satisfied that
\begin{equation*}
G^N(0.5,3.78)=\left\{1,1,0,1,1,0,1,1,1,0,1,1,1,1,1,0,1,1,1,1,0,1,1,1\ldots\right\}.
\end{equation*}
As a result, if we ask the decryption machine to decrypt $(8,0)$,
then we obtain $\left\{1,1,0,1,1,0,1,1\right\}$. Similarly, the
decryption machine will return $\left\{1,0,1,1,1,1,1,0\right\}$ when
the input is $(8,8)$, and $\left\{1,1,1,1,0,1,1,1\right\}$ when the
input is $(8,16)$. In other words, the decryption of the first
ciphertext returns the first $b$ bits of $G^N(x_0,r)$, the
decryption of the second ciphertext gives the second set of $b$ bits
of $G^N(x_0,r)$, and so on.

\subsection{Estimation of the control parameter}
\label{section:parameterEstimation}

If the binary sequence (i.e., the Gray code) derived from the
iteration of the logistic map is known, then it is possible to
infer the value of $r$ based on the concept of Gray Ordering
Number (GON). The GON was introduced in \cite{alvarez98} as a way
to reinterpret the main results of \cite{metropolis73} in a more
intuitive way. The calculation of the GON of a binary sequence
$G^N(x_0,r)$ involves two steps:
    \begin{itemize}
        \item The binary sequence is transformed into another binary
        sequence using the next equation:
            \begin{equation}
                u_i(x_0,r)=\left\{ \begin{array}{ll}g_i(x_0,r), &\textrm{if} \ \ \ i=0, \\
                u_{i-1}(x_0,r)\oplus g_{i}(x_0,r), & \textrm{if} \ \ \ i>0,\end{array}\right.
                \label{eq:binaryForGray}
            \end{equation}
            where $i=\left\{0,1,2,\ldots,N\right\}$.
        \item The GON of the original binary sequence is calculated as:
            \begin{equation}
                GON(G^N(x_0,r))=2^{-1}\cdot u_0+2^{-2}\cdot u_1 + \cdots
                +2^{-N-1}\cdot u_{N}.
            \end{equation}
    \end{itemize}

According to \cite{wu04}, for any concave unimodal map with
critical point equal to $0.5$  it is satisfied that
\begin{equation}
    GON(G^N(f(x_0,r),r)) \leq GON(G^N(f(0.5,r),r))
\end{equation}

for any value of $r$ in $[3,4]$ and any value of $x_0$ in $[0,1]$.
Furthermore, the function $GON(G^N(f(0.5,r),r))$ is an increasing
function with respect to $r$ (see Fig.~3(a) of \cite{wu04}). These
two facts are used in \cite{wu04} to estimate the value of the
control parameter $r$. First of all, the value
$GON(G^n(f(0.5,r),r))$, for $n<N$, is approximated as the maximum
value of the $GON$ of $M$ different shift-left sequences obtained
from $G^N(x_0,r)$. Afterwards, the monotonic relationship between
$GON(G^n(f(0.5,r),r))$ and $r$ is used to obtain an estimation of
$r$ through a binary search procedure.

\begin{figure}[!htbp]
    \centerline{%
    \includegraphics{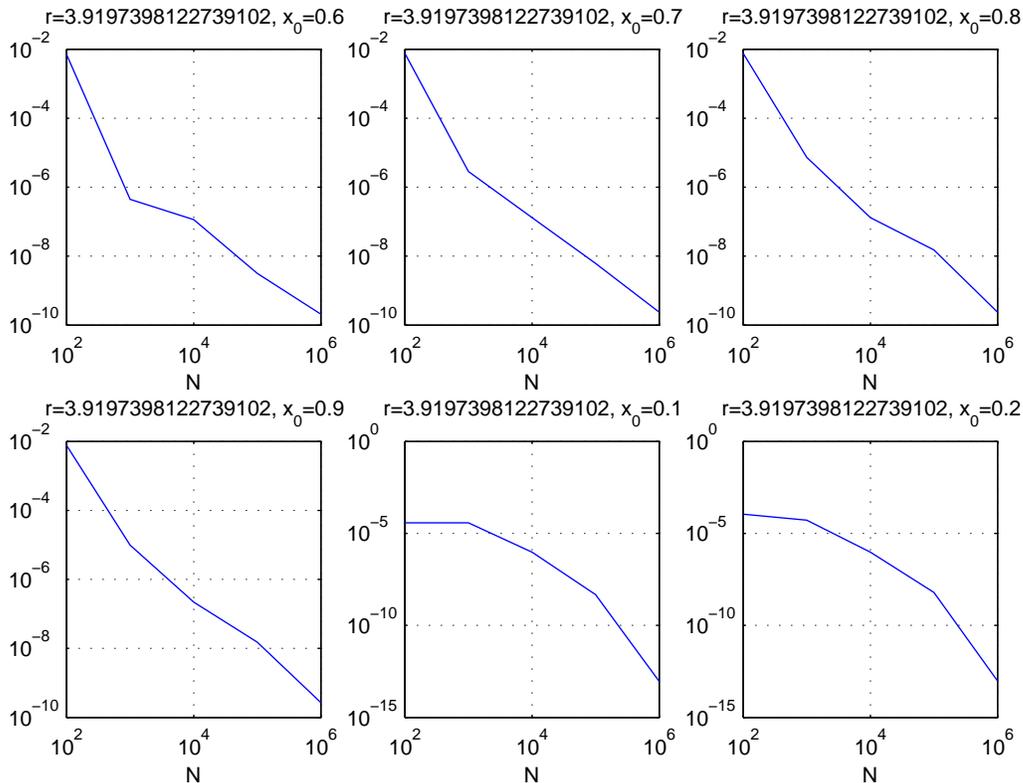}
    }
    \caption{Parameter estimation errors for $r=3.9197398122739102$, different values of $x_0$ and $N$.}
    \label{figure:parameterEstimationError}
\end{figure}

In order to test this algorithm, some simulations have been carried
out. The parameter estimation errors for $r=3.9197398122739102$ are
shown in Fig.~\ref{figure:parameterEstimationError}. Different
values of $x_0$ and $N$ were considered, for a fixed length of the
subsequences of $n=100$. Since this method is based on the
approximation of the maximum of $GON(G^n(f(0.5,r),r))$ through $M$
different values, it is expected that the exact value of $r$ cannot
be obtained unless the value $0.5$ is part of the chaotic orbit from
which the binary sequence was calculated. Moreover, the
characteristic dependency of chaotic maps on the initial condition
makes the parameter estimation error depend on the value of $x_0$,
as shows in Fig.~\ref{figure:parameterEstimationError}.
Nevertheless, the proposed method allows to obtain an estimation of
$r$ which implies a considerable narrowing of the key space and
which can be further improved through a trial and error strategy,
i.e., a brute force attack on the value of the control parameter in
a dramatically reduced key-space.

\subsection{Estimation of the initial condition}
\label{section:x0Estimation}

In this subsection we will assume that we have obtained the exact
value of $r$ by using the algorithm discussed in the last
subsection. Indeed, when considering the security of a cryptosystem,
a partial knowledge of the key must not lead to the determination of
the rest of the key \cite[Rule 7]{Alvarez06a}. Therefore, even if we
were not able to estimate the value of $r$ and obtain the exact
value through a brute-force attack, the recovery of $x_0$ based on
the knowledge of the other subkey $r$ would represent a very
important flaw of the cryptosystem under study.

\begin{figure}[!htbp]
\centerline{%
    \subfigure[$$]{\includegraphics{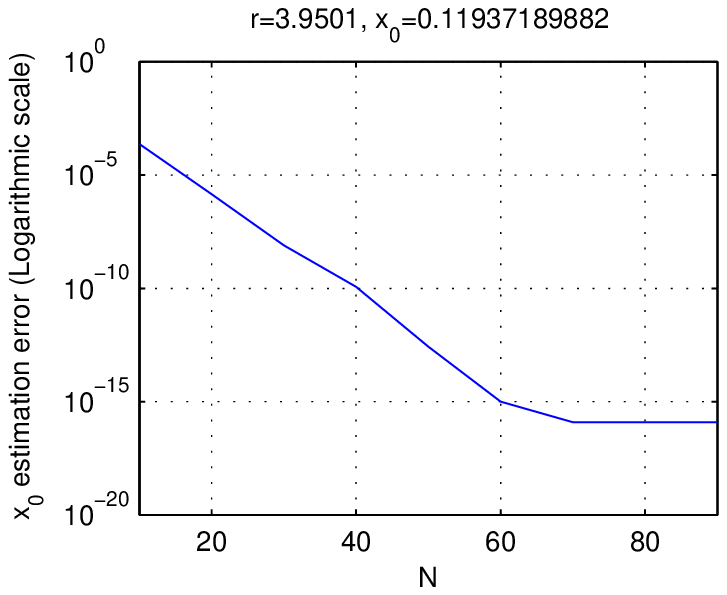}}
    \subfigure[$$]{\includegraphics{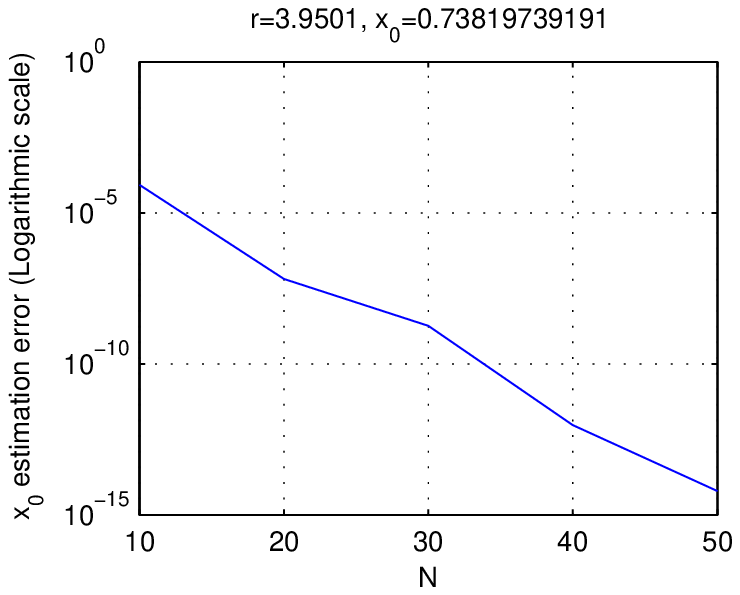}}
    }
\centerline{%
    \subfigure[$$]{\includegraphics{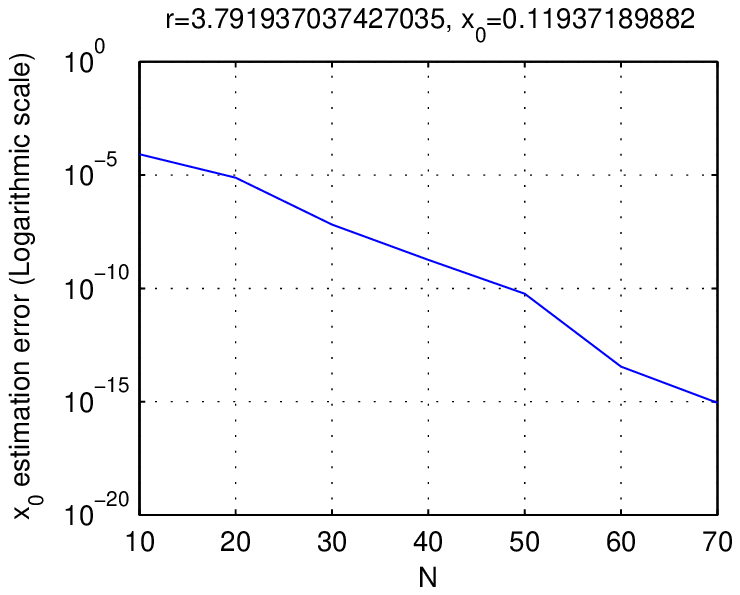}}
    \subfigure[$$]{\includegraphics{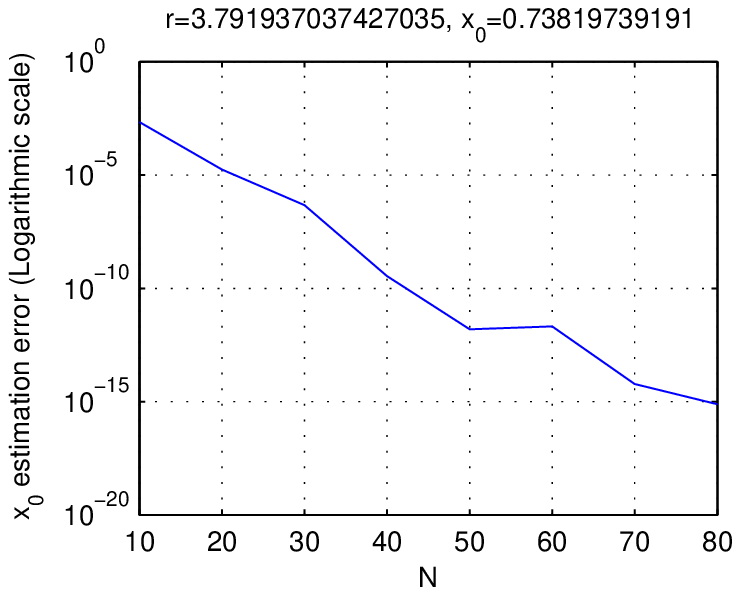}}
} \caption{Initial condition estimation errors for different values
of $r$, $x_0$ and $N$.} \label{figure:x0EstimationError}
\end{figure}

As pointed out in \cite{wu04}, the GON of $G^N(x_0,r)$ is a
monotonic increasing function with respect to $x_0$ (see Fig.~1 of
\cite{wu04}). This means that one can obtain the value of $x_0$
through an iterative algorithm similar to that described in the last
subsection. This algorithm was used to estimate the value of the
initial condition from which $G^N(x_0,r)$ was generated. Different
values of $r$, $x_0$ and $N$ were considered. The results are shown
in Fig.~\ref{figure:x0EstimationError}. For all analyzed situations,
a number of bits greater than $80$ implies an estimation error below
$10^{-15}$. Since all the simulations were performed using double
precision, this means that the exact recovery of the initial
condition is possible.

\section{Other weaknesses}
\label{section:weaknesses}

In this section some other problems of the cryptosystem under
study are emphasized.

\subsection{Considerations about the chaotic system employed}

In \cite{wang08b} it is pointed out that most chaotic systems can be
used to implement the proposed cryptosystem. However, there is no
indication of the requirements that a chaotic system must fulfill to
determine a secure cryptosystem according to the proposed
encryption/decryption structure. Moreover, in the previous section
we proved that at least a family of chaotic maps, i.e., the unimodal
chaotic maps with fixed critical point equal to $0.5$ cannot be used
as long as a high level of security against chosen-ciphertext attack
is needed. Furthermore, a different way should be used to generate
the binary sequence for the encryption procedure. In the original
design, this binary sequence is obtained by comparing each chaotic
state included in a chaotic orbit with the fixed threshold value
0.5. Nevertheless, to ensure good statistical characteristics of the
binary sequence, the threshold value should be selected according to
the dynamics of the underlying chaotic system.

\subsection{Considerations about the chaotic orbit generation}

The characteristics of a cryptosystem should be precisely defined in
order to facilitate its implementation \cite[Rule 1]{Alvarez06a}.
During the encryption step of the cryptosystem under consideration,
the plaintext is divided into a set of binary sequences $P_i$ which
are successively located in the binary sequence $G^N(x_0,r)$. It is
possible that $P_i$ is not included in $G^N(x_0,r)$. In this case,
the length of $P_i$ is progressively decreased until it is found in
$G^N(x_0,r)$. Nevertheless, there is no information about the length
of $G^N(x_0,r)$, i.e., about the maximum number of iterations
$N_{max}$ needed to conclude whether the length of $P_i$ must be
decreased. Furthermore, not only the length of $G^N(x_0,r)$ is not
explicitly established, but also some interpretation problems
concerning the precise way of generating $G^N(x_0,r)$ can be found.
Fist of all, in \cite{wang08b} it is not mentioned whether the first
bit of $G^N(x_0,r)$ corresponds to $x_0$ or to $x_1$. On the other
hand, once the plain block $P_i$ has been encrypted, it is not clear
whether the next binary sequence starts from $G^N(x_{n_i},r)$,
$G^N(x_{n_i+1},r)$ or $G^N(x_{n_i+b},r)$. Note that we fixed these
problems in our description of the cryptosystem given in
Sec.~\ref{section:description}.

\subsection{Considerations about the key space}

\begin{figure}[!htbp]
    \centering
    \includegraphics{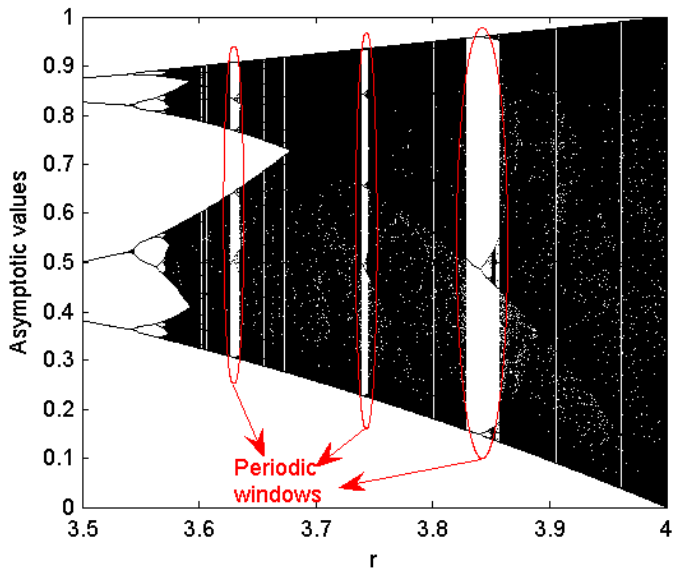}
    \caption{Bifurcation diagram of the logistic map showing the existence of periodic windows.}
    \label{figure:logBif}
\end{figure}

The inadequacy of the logistic map for the implementation of this
cryptosystem has been proved by means of a ciphertext attack.
However, the selection of this map entails another important problem
that suggests  not to choose the logistic map as a base of any
cryptosytem \cite{arroyo08b}. This problem concerns the definition
of the key space. In \cite{wang08b} it is claimed that the value of
the control parameter $r$ should be selected within the interval
$(3.57148....,4)$ to exhibit a chaotic behavior. However, the
existence of periodic windows in this region is well known (see
Fig.~\ref{figure:logBif}) and so the selection of $r$ should be
performed in a more precise manner in order to avoid these
\cite[Rule 5]{Alvarez06a}.

\section{Conclusions}
\label{section:conclusions}

Some weaknesses of the chaotic cryptosystem described in
\cite{wang08b} have been discussed in this paper. A
chosen-ciphertext attack has been described, which can recover the
secret key of the cryptosystem by exploiting the theory of symbolic
dynamics. Some other problems related to the design of the
cryptosystem have also been pointed out. As a result, we recommend
not to use this algorithm for secure applications.

\section{Acknowledgments}

The work described in this paper was supported by \textit{Ministerio
de Educaci\'on y Ciencia of Spain}, research grant SEG2004-02418,
\textit{Ministerio de Ciencia y Tecnolog\'{i}a} of Spain, research
grant TSI2007-62657 and \textit{CDTI, Ministerio de Industria,
Turismo y Comercio of Spain} in collaboration with Telef\'onica I+D,
Project SEGUR@ with reference CENIT-2007 2004. Shujun Li was
supported by a research fellowship from the \textit{Alexander von
Humboldt Foundation, Germany}.

\bibliographystyle{elsart-num}

\begin{thebibliography}{10}
\expandafter\ifx\csname url\endcsname\relax
  \def\url#1{\texttt{#1}}\fi
\expandafter\ifx\csname urlprefix\endcsname\relax\def\urlprefix{URL
}\fi

\bibitem{alvarez99}
E.~Alvarez, A.~Fern\'{a}ndez, P.~Garc\'{\i}a, J.~Jim\'{e}nez,
A.~Marcano, New
  approach to chaotic encryption, Physic Letters A 263 (1999) 373--375.

\bibitem{alvarez00}
G.~Alvarez, F.~Montoya, M.~Romera, G.~Pastor, Cryptanalysis of a
chaotic
  encryption system, Physics Letters A 276 (2000) 191--196.

\bibitem{li2001}
S.~Li, X.~Mou, Y.~Cai, Improving security of a chaotic encryption
approach,
  Physics Letters A 290~(3-4) (2001) 127--133.

\bibitem{wang08b}
X.~Wang, C.~Duan, N.~Gu, A new chaotic cryptography based on
ergodicity,
  International Journal of Modern Physics B 22~(7) (2008) 901--908.

\bibitem{alvarez98}
G.~Alvarez, M.~Romera, G.~Pastor, F.~Montoya, Gray codes and {1D}
quadratic
  maps, Electronic Letters 34~(13) (1998) 1304--1306.

\bibitem{cusick99}
T.~Cusick, Gray codes and the symbolic dynamics of quadratic maps,
Electronic
  Letters 35~(6) (1999) 468--469.

\bibitem{alvarez03a}
G.~Alvarez, F.~Montoya, M.~Romera, G.~Pastor, Cryptanalysis of an
ergodic
  chaotic cipher, Physics Letters A 311 (2003) 172--179.

\bibitem{wu04}
X.~Wu, H.~Hu, B.~Zhang, Parameter estimation only from the symbolic
sequences
  generated by chaos system, Chaos, Solitons and Fractals 22 (2004) 359--366.

\bibitem{physcon07}
G.~Alvarez, D.~Arroyo, J.~Nunez, Application of gray code to the
cryptanalysis
  of chaotic cryptosystems, in: 3rd International IEEE Scientific Conference on
  Physics and Control (PhysCon'2007, 3rd - 7th, September 2007, Potsdam,
  Germany), IEEE IPACS, Potsdam, Germany, 2007.
\newline\urlprefix\url{http://lib.physcon.ru/?item=1355}

\bibitem{stinson95}
D.~Stinson, Cryptography: Theory and Practice, CRC Press, 1995.

\bibitem{metropolis73}
N.~Metropolis, M.~Stein, P.~Stein, On the limit sets for
transformations on the
  unit interval, Journal of Combinatorial Theory, Series A 15~(1) (1973)
  25--44.

\bibitem{Alvarez06a}
G.~Alvarez, S.~Li, Some basic cryptographic requirements for
chaos-based
  cryptosystems, International Journal of Bifurcation and Chaos 16~(8) (2006)
  2129--2151.

\bibitem{arroyo08b}
D.~Arroyo, G.~Alvarez, V.~Fernandez, On the inadequacy of the
logistic map for
  cryptographic applications, arXiv:0805.4355 (2008).
\newline\urlprefix\url{http://arxiv.org/abs/0805.4355}

\end{thebibliography}

\end{document}